\documentclass[12pt,preprint]{aastex}

\usepackage{lscape}
\newcommand{\feh} {\mbox{\rm [Fe/H]}}
\newcommand{\teff} {\mbox{\rm $T_{\rm{eff}}$}}
\newcommand{\logg} {\mbox{{\rm log(g)}}}

\begin{document}


\title{Absolute Magnitudes of Seismic Red Clumps in the {\it Kepler} Field and
  SAGA: the age dependency of the distance scale}

\author{Y.Q. Chen$^{1,2,3}$, L. Casagrande$^{4}$, G. Zhao$^{1,2}$,  J. Bovy$^3$, V. Silva Aguirre
$^5$, J.K. Zhao$^1$ and Y.P. Jia$^{1,2}$}
\altaffiltext{1}{Key Laboratory of Optical Astronomy, National Astronomical
Observatories, Chinese Academy of Sciences, A20 Datun Rd, Chaoyang District, Beijing, 100012, China;
cyq@bao.ac.cn.}
\altaffiltext{2}{University of Chinese Academy of Sciences, 19 A Yuquan Rd, Shijingshan District, Beijing, P.R.China 100049}
\altaffiltext{3}{Astronomy \& Astrophysics Department, University of Toronto, 50 St. George Street, ON, Canada}
\altaffiltext{4}{Research School of Astronomy \& Astrophysics, Mount Stromlo Observatory, The Australian National University, ACT 2611, Australia}
\altaffiltext{5}{Stellar Astrophysics Centre, Department of Physics and Astronomy, Aarhus University,Ny Munkegade 120, DK-8000 Aarhus C, Denmark}

\begin{abstract}
Red clump stars are fundamental distance indicators in astrophysics, although theoretical stellar models predict a dependence of absolute magnitudes with ages. This effect is particularly strong below $\sim 2$ Gyr, but even above this limit a mild age dependence is still expected.
We use seismically identified red clump stars
in the {\it Kepler} field for which we have reliable distances, masses and
ages from the SAGA survey to first explore this effect. By excluding red clump
stars with masses larger than $1.6 M_{\odot}$ (corresponding to ages younger
than 2 Gyr), we derive robust calibrations linking intrinsic colors to
absolute magnitudes in the following photometric systems: Str\"omgren $by$,
Johnson $BV$, Sloan $griz$, 2MASS $JHK_s$ and WISE $W1W2W3$. With the precision
achieved we also detect a slope of absolute magnitudes
$\sim 0.020\pm0.003\, \rm{mag}\,\rm{Gyr}^{-1}$ in the infrared, implying that
distance calibrations of clump
stars can be off by up to $\sim0.2\, \rm{mag}$ in the infrared (over the
  range from 2 Gyr to 12 Gyr) if their ages are unknown. Even larger
uncertainties affect optical bands, because of the stronger interdependency of
absolute magnitudes on colors and age. Our distance calibrations are ultimately
based on
asteroseismology, and we show how the distance scale can be used to test the
accuracy of seismic scaling relations. Within the uncertainties our
calibrations are in agreement with those built upon local red clump with
{\it Hipparcos} parallaxes, although we find a tension which if confirmed
would imply that scaling relations overestimate radii of red clump stars
by $2\pm2$\%. Data-releases post {\it Gaia} DR1 will provide
an important testbed for our results.

\end{abstract}

\section{Introduction}
The clump of red giant stars is a ubiquitous feature in (nearly) equidistant
stellar populations.
Theoretically predicted by \cite{Thomas67} and \cite{Iben68} it was first
recognized in the color--magnitude diagram of old- and intermediate-age open
clusters \citep{Cannon70}, and later also
observed in ``metal-rich'' globular clusters \citep[e.g.][]{HH},
towards the Galactic bulge \citep[e.g.][]{Pac} and in nearby galaxies
\citep[e.g.][]{Stanek98}. Red Clump stars (hereafter RCs) also constitute a
quite remarkable and well populated feature in the color magnitude diagram of
nearby field stars once precise distances from {\it Hipparcos} are used
\citep[e.g.][]{Pac,Girardi98}.
It is well established that RCs have nearly constant absolute magnitudes,
and once identified (e.g.\,as an overdensity in a equidistant stellar
population) they are important standard candles for deriving distances.

The identification of RCs among field stars has been difficult so far, due to
the limited number of them with precise trigonometric parallaxes, combined to
the fact that the
{\it Hipparcos} ``sphere'' covers a rather limited volume, extending to
distances of order 100~pc. While {\it Gaia} is due to shift this limit to
several kpc \citep{Lindegren16}, space-borne asteroseismic missions such as
CoRoT \citep{Auvergne09} and {\it Kepler} \citep{Gilli} already allow us to
derive
stellar distances for stars with measured solar-like oscillations, among which
are RCs \citep[e.g.][]{Silva12,Miglio13,C14a,Rod14}. Further when
period-spacing information is available, asteroseismology is also able to
unambiguously distinguish between stars ascending the red giant branch (RGB)
burning hydrogen in a shell, and those (i.e. RCs) that have already ignited
helium burning in their cores \citep[e.g.][]{Montalban10,Bedding11,Stello13}.

In this work, we aim at deriving color and absolute magnitude calibrations
in many photometric systems for seismically-identified RCs and compare these
calibrations with those available in the literature for local RCs.
Our goals are manifold. First, we aim at obtaining a more reliable
selection of RCs compared to other studies appeared in the literature so far:
taking advantage of the seismic period-spacing we can in fact precisely
identify {\it bona-fide} RCs. This allows us to remove from our sample
contaminants (among which are stars going through the bump in the red giant
branch), which instead plague other
RCs selection techniques. Second, we take advantage of seismic distances to
derive reliable absolute magnitudes for all our RCs. Seismic distances are
obtained scaling stellar angular diameters \citep[in our case determined from
the InfraRed Flux Method, see][]{C14a} to seismic radii \citep[ultimately
based on scaling relations, see e.g.,][]{Stello09,Miglio09}.
A good deal of efforts is currently invested to test the accuracy of scaling
relations, and whether they have any dependence on other parameters such as
e.g., metallicity and evolutionary phase
\citep{White11}. Radii derived from scaling relations have been shown to be
accurate to about 5\%, depending on evolutionary status \citep[e.g.,][]
{Huber12,Silva12,White13,Gaulme16}, although they are considerably less
tested in the RC regime \citep[for a summary see e.g.,][]{Miglio13b,Bro16}.
Currently, uncertainty on seismic radii is one of the limiting factor in the
accuracy at which seismic stellar distances can be derived. The other stems from
the accuracy at which stellar effective temperatures (and thus angular
diameters) can be derived from photometry \citep{C14b}. Both sources of
uncertainty however are distance independent (modulo reddening), meaning that
for a distance fractional error $f$, seismic distances will be superior to
astrometric ones beyond $10^{6} f/\omega$ parsec (where $\omega$ is the
parallax error in $\mu{\rm{as}}$). Here we use seismic distances from
\cite{C14a}, which have a median uncertainty of 3.3\% (assuming no
systematic errors in the adopted scaling relations) and typical distances above
1~kpc. {\it Gaia} DR1 parallaxes have a systematic error of
$300\mu{\rm{as}}$ in addition to random errors \citep{Lindegren16},
effectively meaning that our seismic distances are always more precise than
{\it Gaia} DR1.
Finally, from seismology we also know masses and ages of our RC stars,
meaning that we can investigate the dependence of absolute magnitudes on these
parameters, important to assess the range within which RC absolute magnitude
calibrations can be trusted.

\section{The Red Clump sample}

The identification of RCs has been traditionally carried out by eye, selecting
stars in the HR diagram having a location consistent with their presence.
Despite RCs have very different internal structure from stars ascending the
RGB, a clean selection between the two has been impossible so far, since they
occupy nearly the same position in luminosity, effective temperature, gravity
and colors within the observational uncertainties. Asteroseismology has
recently allowed to overcome this limitation, since in the $\Delta \nu$ versus
$\Delta P$ diagram (here $\Delta \nu$ is the frequency shift of consecutive
overtone modes of the same degree, and $\Delta P$ is the pairwise period
spacing between adjacent dipole modes), RCs are
clearly separated from red giants \citep[e.g.,][]{Stello13}.
For the purpose of our work we want a sample of seismically identify RCs,
which also has information on their metallicities, radii, distances, masses,
ages as well as magnitudes in various photometric systems. This is possible
thanks to the Str\"omgren survey for Asteroseismology and Galactic
Archaeology \citep[SAGA,][]{C14a,C16}. Here we use seismic ages derived
assuming no mass-loss: this is motivated by the fact that recent studies seem
to indicate a low efficiency of mass-loss \citep[see discussion in ][]{C16}.

Fig.~1 shows the $\teff - \logg$ plane for the entire SAGA sample; an
overdensity of stars is present for $\logg \sim 2.5 \pm 0.1$~dex, but solely
based on this information is impossible to single out RC stars. In SAGA, a
large fraction of the object with seismic information also has evolutionary
phase classification, based on period spacing \citep{Stello13}. The latter
tells us whether a star is evolving along the RGB with a hydrogen burning shell
or already in the RC phase. We thus use this information to limit our sample
to objects marked as RC (green open circles in Fig.~1). Overplotted with crosses
are also seismically inferred members of the open cluster NGC\,6819, some of
which are RCs as well.

RCs with masses above $\gtrsim 1.8 M_\sun$ ignite helium in nondegenerate
conditions which observationally results in slightly fainter luminosities and
hotter effective temperatures \citep[e.g.,][]{Girardi99}. This feature is known
as secondary clump and it is clearly visible with our data \citep[see Fig.~1,
and also discussion in][]{C14a}. Clearly, the absolute magnitude of secondary
clump stars deviate significantly from a constant value, which in turn
prevents them to be used as good distance calibrators. Because of this,
firstly, we
exclude from our sample all RCs with masses above $1.6 M_\sun$ and
ages younger thant 2~Gyr. In fact, the age of stars in the red giant phase,
either RGB or RC, is largely determined by the time spent in the main
sequence core-hydrogen burning phase, thus meaning that the mass of a red
giant is also a good proxy for its age.
The conversion between mass and age introduces a dependency on stellar
models, which among other things is sensitive e.g., on overshooting during
the main sequence, and mass-loss. For stars along the red giant branch, mass-loss mostly occurs towards the tip of the RGB and the clump phase, and thus it could potentially impact the age determination of our stars. However, recent studies suggest that mass-loss is rather inefficient (see discussion in Casagrande et al. 2016), meaning that it only moderately affects our ages. All our ages are derived assuming no mass-loss (Reimers' parameter $\eta=0$), but in Section 3.7 we discuss how our results would change in the case of an extremely efficient mass-loss ($\eta=0.4$, which however is currently disfavoured by observations).
 Overshooting during main sequence for
$M < 1.6 M_\sun$ stars does not change the mass of the degenerate He core,
which follows the pattern of $\Delta P$ (an important parameters for mass
determination) according to \cite{Montalban13}.
With the above-mentioned  selection procedure, all RCs in our sample have a
very narrow range
of surface gravities $2.3 \lesssim \logg \lesssim 2.6$~dex, while their
effective temperatures vary between $\sim 4500$~K to $5200$~K. Secondly,
we exclude members of NGC6819 because they are mostly secondary clump being a
young cluster.  We also exclude KIC6206407, a RC star with a second oscillation
signal in the Kepler data, indicating a likely binary \citep{C14a}.
Finally, we also exclude all stars with bad metallicity flag in
SAGA, i.e. keep stars with $Mflg\,=\,0$ only, for a
final sample of 171 stars. This is the sample of RCs that will be used in the
following of the paper to derive our color--absolute magnitude relations in
different photometric system. Further pruning of the sample to retaining only
stars with best photometric measurements in a given system will be done as
described in the next Section.

\begin{figure*}
\epsscale{1.0}
\plotone{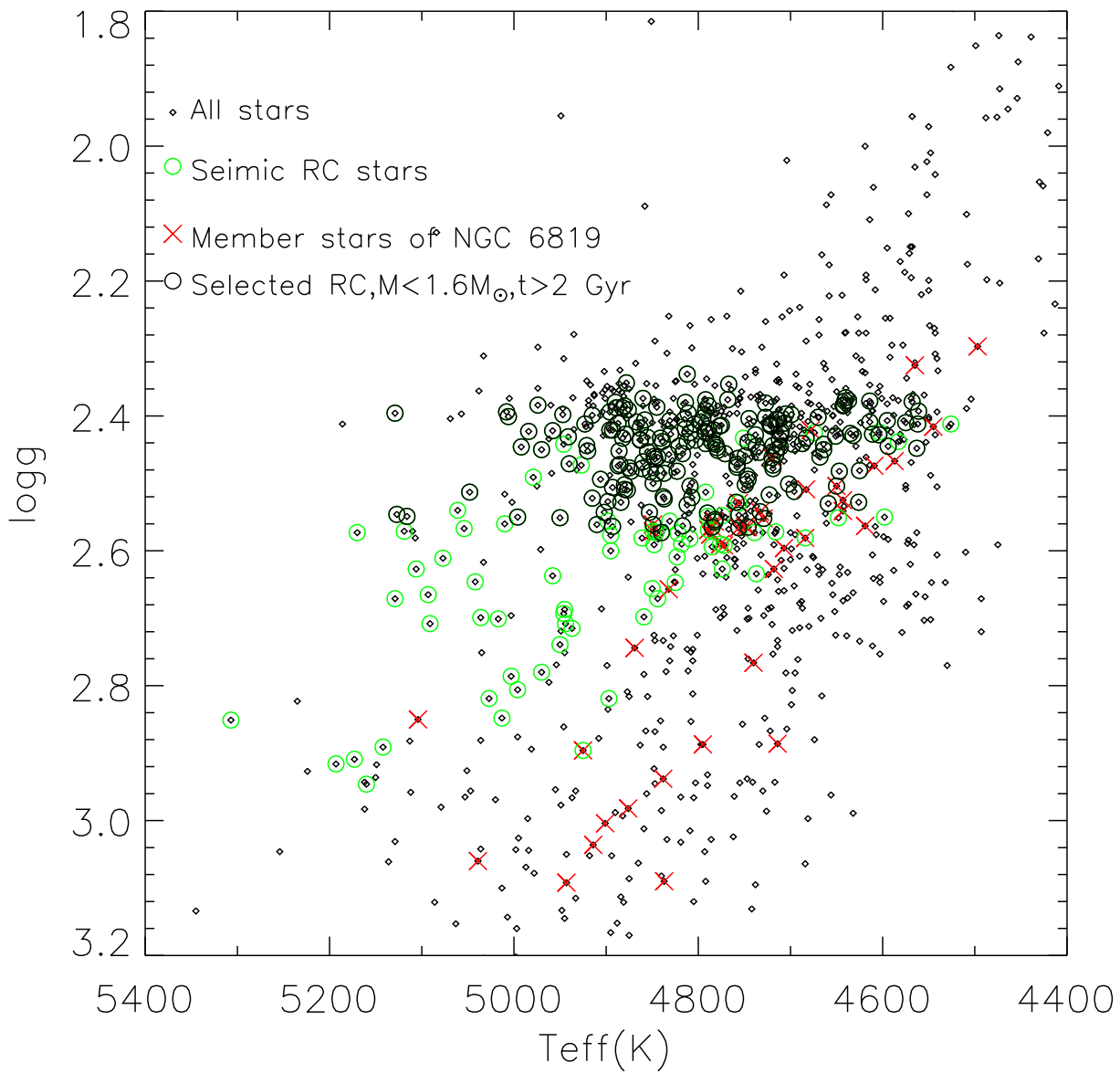}
\caption{$\teff$ versus $\logg$ diagrams for all stars in SAGA (open diamonds),
highlighting stars seismically identified as RC (green open circles), members
of the young open cluster NGC\,6819 (red crosses), and seismically identified
RCs with $M<1.6$ $M_\sun$ and age\,$>2$\,Gyr (black open circles).}
\end{figure*}

\section{Color and magnitude calibrations in different photometric systems}

In addition to our Str\"omgren observations, magnitudes in the following
photometric systems are also available for most of the targets:
$BV$ and $g'r'i'$ from APASS \citep{Henden09}, $griz$ from
the Kepler Input Catalog \citep[KIC,][]{Brown11}, $JHK_s$ from 2MASS
\citep{Cutri03} and $W1W2W3$ from WISE \citep{Wright10}.
All seismic targets have apparent magnitudes in the range
$10 \lesssim V \lesssim 14$, meaning that photometric errors are usually small
in all of the above systems, with typical uncertainties varying
between $0.01$ and $0.03$~mag. We discuss each photometric system and
the quality cuts adopted on the photometry in the following sub-sections.

Before doing this though, reddening must be properly taken into account to
derive correct intrinsic colors and absolute magnitudes.
SAGA provides reddening $E(B-V)$ for all asteroseismic targets. We use
these values to deredden all photometric measurements, using extinction
coefficients appropriate for clump stars. These are computed as described in
\cite{CV}; briefly, we apply the \cite{ccm89} extinction law to a synthetic
spectrum representative of clump stars ($\teff=4750\,K$, $logg=2.0$ and
$\feh=0$) from which the following coefficients are derived: $R_B=3.934$ and
$R_V=3.086$ for the Johnson system, $R_g=3.669$, $R_r=2.687$, $R_i=2.106$ and
$R_z=1.517$ for the Sloan $griz$ system, $R_b=3.846$ and $R_y=3.124$ for the
Str\"omgren system, $R_J=0.894$, $R_H=0.566$ and $R_{K_{S}}=0.366$ for the
2MASS system, and $R_{W1}=0.242$, $R_{W2}=0.134$ and $R_{W3}=0.349$ for the
WISE system. These values are consistent with those published in the
literature, but have the advantage of being computed using a reference
spectrum appropriate for RCs, thus ensuring better consistency among different
bands.

Once color excess and extinction coefficients are know, dereddened magnitudes
in any given band $\zeta$ can be derived as $m_{\zeta,0} = m_\zeta - R_\zeta
E(B-V)$, from which dereddened absolute
magnitudes $M_{\zeta,0} = m_{\zeta,0} - 5\log D+5$,
where $D$ is the distance (in parsec), also determined from SAGA\footnote{Note
  that in the rest of the paper and figures, all colors and absolute magnitude
  are always corrected for reddening, even though to make the notation more
  readable the subscript $0$ is not included.}.
Distances in SAGA are obtained scaling angular diameters computed via the
InfraRed Flux Method to asteroseismic
radii. The distance of each star is then fed into empirically calibrated,
three dimensional Galactic extinction models to
derive its reddening, with an iterative procedure to converge in both distance
and reddening.
Individual uncertainties on reddening are not available, but those are expected
to be of order $\pm 0.02$~mag on average \citep{C14a}, given that the SAGA
sample used here
covers a stripe with Galactic latitude between $8^{\circ}$ and $20^{\circ}$, where
reddening is relatively low. Thus, the average colors of stars vary by
considerable less than this uncertainty. Further, the zero point of our
reddening values is anchored to the open cluster NGC\,6819 for which a robust
value of $E(B-V)=0.14 \pm 0.01$ is available from the literature. On average,
reddening uncertainties are thus at the level of few hundredths of a magnitude,
having negligible impact on the color--absolute magnitude calibrations
presented later in the paper.

As a further check on the reddening values adopted from SAGA, we also derive
independent estimates using the relation
$E(B-V)= A_{K_s}/0.366=0.918(H-W2-0.08)/0.366$, where the expression for
$A_{K_s}$ is based on the Rayleigh-Jeans Color Excess method \citep[RJCE,][]
{Majewski11}. This technique relies on the near-constancy of the
infrared color $H-$[4.5$\mu$m]) for evolved stars. In our case we adopt $H$ and
$W2$ magnitudes from 2MASS and WISE, respectively. The latter filter is
centred on a wavelength of 4.6$\mu$m which is taken into account by the
$-0.08$ factor in $A_{K_s}$ as reported in \cite{Majewski11}.

As an example of the different precision achieved with different sets of
reddening values, Fig.~2 shows the $(b-y)$ vs.~$M_{y}$ relation for RCs when
adopting $E(B-V)$ from SAGA or the RJCE calibration instead. In the
latter case, the scatter of the color--absolute magnitude relation is
considerably larger (0.14 vs. 0.10). In particular, stars belonging to the
open cluster
NGC\,6819 cover a broad range of colors, whereas when switching to the
reddening values from SAGA it narrows around $(b-y) \simeq 0.63$.

In this paper linear fits are obtained using the IDL linear regression
routine \emph{regress}. Uncertainties stemming from photometric and distance
errors are taken into account in doing the fits. For each fit we also compute
the Pearson and Spearman coefficients to measure the strength of correlations.
Results from this method are labeled as \emph{IDL-reg} in what follows. To
check the robustness of our results to potential outliers, we also employ two
additional methods. One method models outliers using a mixture model consisting
of a straight line, mixed with a broad Gaussian to capture outliers. We adapted
the Python code \emph{exMix1} of \cite{Hogg10} and label the results from this
method as \emph{Py-exMix1}. The other method performs a robust linear
regression using an M estimator, by employing iterated re-weighted least
squares, as implemented in the \emph{R} function \emph{rlm}; we label results
using this method as \emph{R-rlm}.

\begin{figure*}
\epsscale{0.8}
\plotone{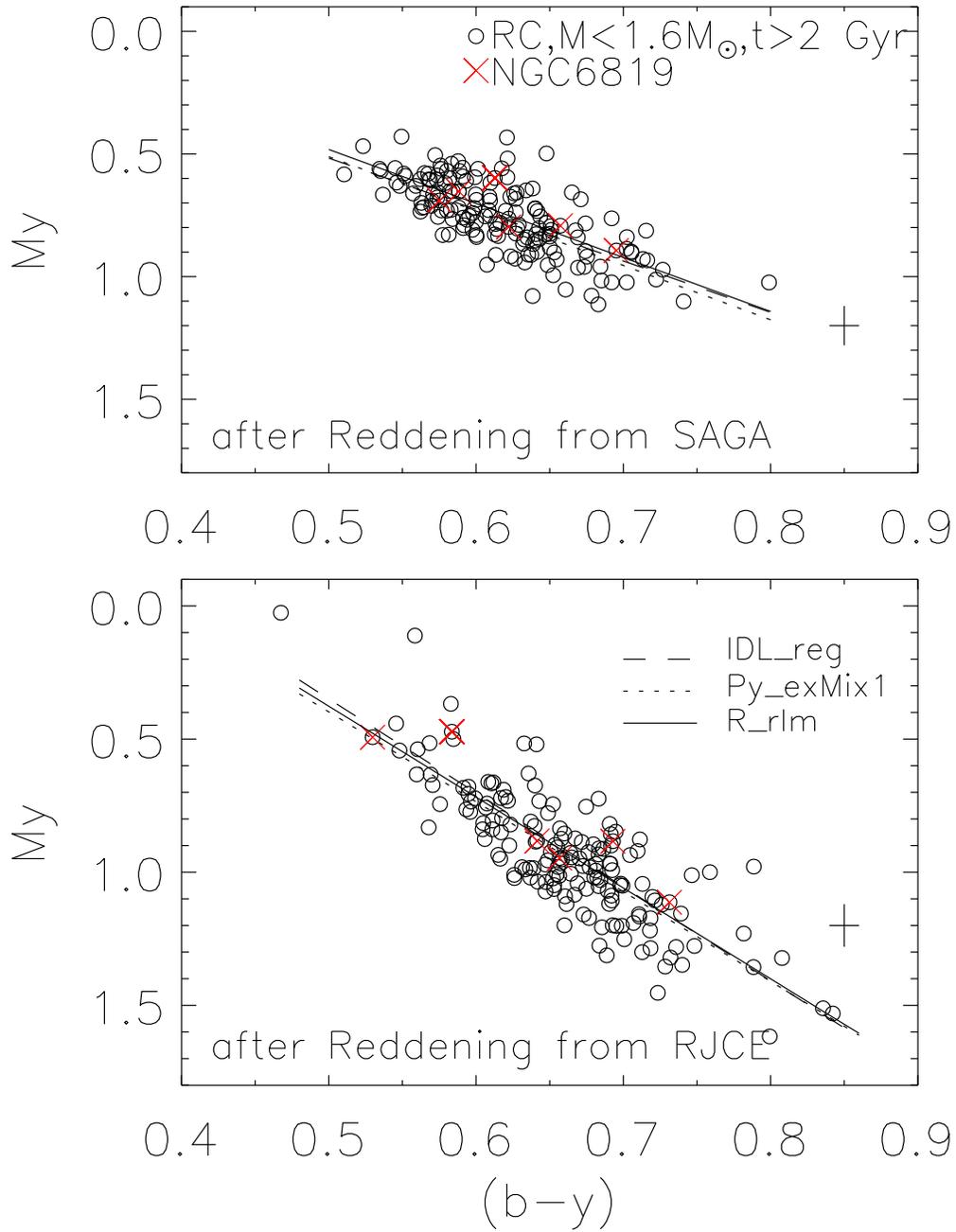}
\caption{The $(b-y)$ versus $M_{y}$ diagrams for RC stars based on reddening
from the SAGA (top panel) and the RJCE method (bottom panel). Seismically
identified members of the open cluster NGC\,6819 are shown by crosses.
Stars with photometric errors in either $b$ or $y$ larger than $0.03$\,mag are
excluded. Linear fits based on three methods (see description in the text)
are shown by dash (\emph{IDL-reg}), dot (\emph{Py-exMix1}) and solid
(\emph{R-rlm}) lines, respectively. All colors and magnitudes are corrected for
reddening.}
\end{figure*}

\subsection{Str\"omgren}

As shown in Fig.~2, the bulk of RCs cover the color range $0.50 \lesssim (b-y)
\lesssim 0.75$\,mag and have absolute magnitudes between
$0.4 \lesssim M_{y} \lesssim 1.1$ mag. Clearly, $M_{y}$ has a steep dependence
on $(b-y)$ color, which we linearly fit ({\it IDL-reg}) obtaining the
following relation $M_{y}=2.209 (b-y)-0.626$ ($\sigma=0.10$)
using 162 RCs with good quality data. Note that stars with photometric errors
in either $b$ or $y$ larger than $0.03$\,mag are
excluded here.
The relations of $M_{y}=2.206 (b-y)-0.620$ from {\it R-rlm} method
and of  $M_{y}=2.220 (b-y)-0.613$ from {\it Py-exMix1} are very close to
the {\it IDL-reg} results, and agree within the scatter. Both the Pearson
and Spearman coefficients are 0.72, indicating a strong correlation.

When the metallicity is taken into account, we find
$M_{y}=2.097 (b-y)+ 0.045 [Fe/H]-0.532$ ($\sigma=0.102$) based on {\it IDL-reg}.
Generally, RC stars in our sample have a metallicity range of $-0.5<\feh<0.5$,
which corresponds to a variation of $\lesssim 0.04\,$mag in $M_{y}$. This
variation is smaller than the scatter of the $M_{y}$ calibration. Thus, in
the following of the work we only provide calibrations linking absolute
magnitudes to colors.
We also explored whether introducing a second order term in color improved
upon the residual of our fit, but find this not to be the case. In fact,
the major source of uncertainty in our calibration is represented by a mean
uncertainty of $0.08\,$mag in absolute magnitudes as a consequence of our
typical distance uncertainties.

\subsection{Johnson}

Fig.~3 shows the $(B-V)$ versus $M_{V}$ diagram, and
the absolute magnitude distributions in Johnson system for RC stars.
In the Johnson $BV$ system, stars with APASS measurement uncertainties larger
than $0.05\,$mag in either $B$ or $V$ band are removed. Most RCs are located
in the range $0.6 \lesssim (B-V) \lesssim 1.3\,$mag and
$0.5 \lesssim M_{V} \lesssim 1.0\,$mag. Based on 119 stars, the scatter is
$0.15\,$mag. A more strict limit on the measured errors of less than
$0.03\,$mag in either B or V band does not improve the correlation and the
star number of the sample is reduced to 53 stars. This scatter likely reflects
the quality of the APASS magnitudes (see also next Section). Note that
fits to the color and absolute magnitude relation based on the three
methods,\emph{IDL-reg}, \emph{Py-exMix1} and \emph{R-rlm} are quite similar.
The Pearson (Spearman) correlation is 0.13 (0.18), indicating a weak
correlation between absolute magnitude and color, as already apparent from
Fig.~3.

\begin{figure}
\epsscale{1.0}
\plotone{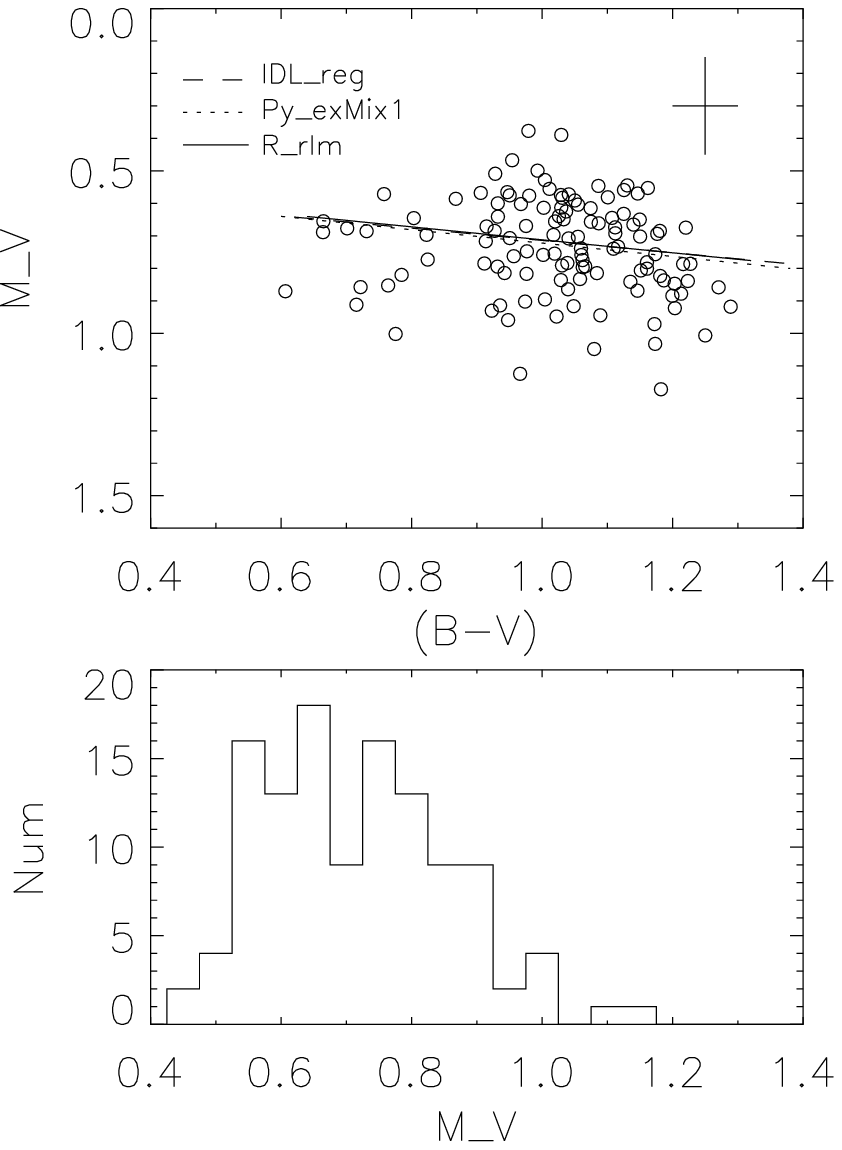}
\caption{The $(B-V)$ versus $M_{V}$ and the color and absolute
  magnitude distributions in Johnson system, for our sample stars. All colors
  and magnitudes are corrected for reddening.}
\end{figure}

\subsection{Sloan}

Magnitudes in the $griz$ system are available from the Kepler Input Catalog,
with typical
uncertainties of $0.02$ mag \citep{Brown11}. However, KIC magnitudes are not
exactly on the Sloan system, and thus have been corrected with the
transformations provided by \cite{Pinsonneault12}.
In addition, for a large fraction of stars $g'r'i'$ magnitudes are also
available from the APASS survey; these are defined in the {\it primed} system
and thus have been converted into the Sloan system using the
transformations of \cite{Tucker06}. However, we also note that in the
color--absolute magnitude plane the scatter is larger when using APASS
magnitudes instead of KIC, thus pointing to lower precision for the former
measurements. Therefore, in the following of the analysis we will use only
KIC $griz$ magnitudes.

Fig.~4 shows the color--absolute magnitude diagrams in different bands. A few
stars are marked with red crosses, and removed from the rest of the analysis:
they are somewhat offset from the bulk of other stars, and have been
identified as anomalous from their 2MASS colors (see next Section).
Absolute magnitudes in each $griz$ band vary linearly with colors, and their
slopes flatten moving to filter centered at longer wavelengths, i.e. from
$M_{g}$ to $M_{z}$. Two stars with $(g-r)>1.0$ seem to deviate from the
linear trend of $M_{g}$ versus $(g-r)$. The number of points is
too little to draw further conclusions, however we advise caution from using
the the calibration at $(g-r)>1.0$. Panels in Fig.~4 display a correlation
between colors and absolute magnitudes which vary depending on the filter, and
decreases as moving to redder filters (see Table 1 for a list of Pearson
and Spearman coefficients).
We also note that the decrease of $M_{i}$ with $(r-i)$ is
consistent with the result of \cite{Zhao01} who found a dependence of Johnson
$M_I$ with $(V-I)$. \cite{Chen09} suggested that $i$ and $z$ are the best
bands for distance calibration of red clump/red horizontal branch stars in the
Sloan system. Here we find that $M_{g}$ versus $(g-r)$ provides an
equally good distance calibration.

\begin{figure*}
\epsscale{1.0}
\plotone{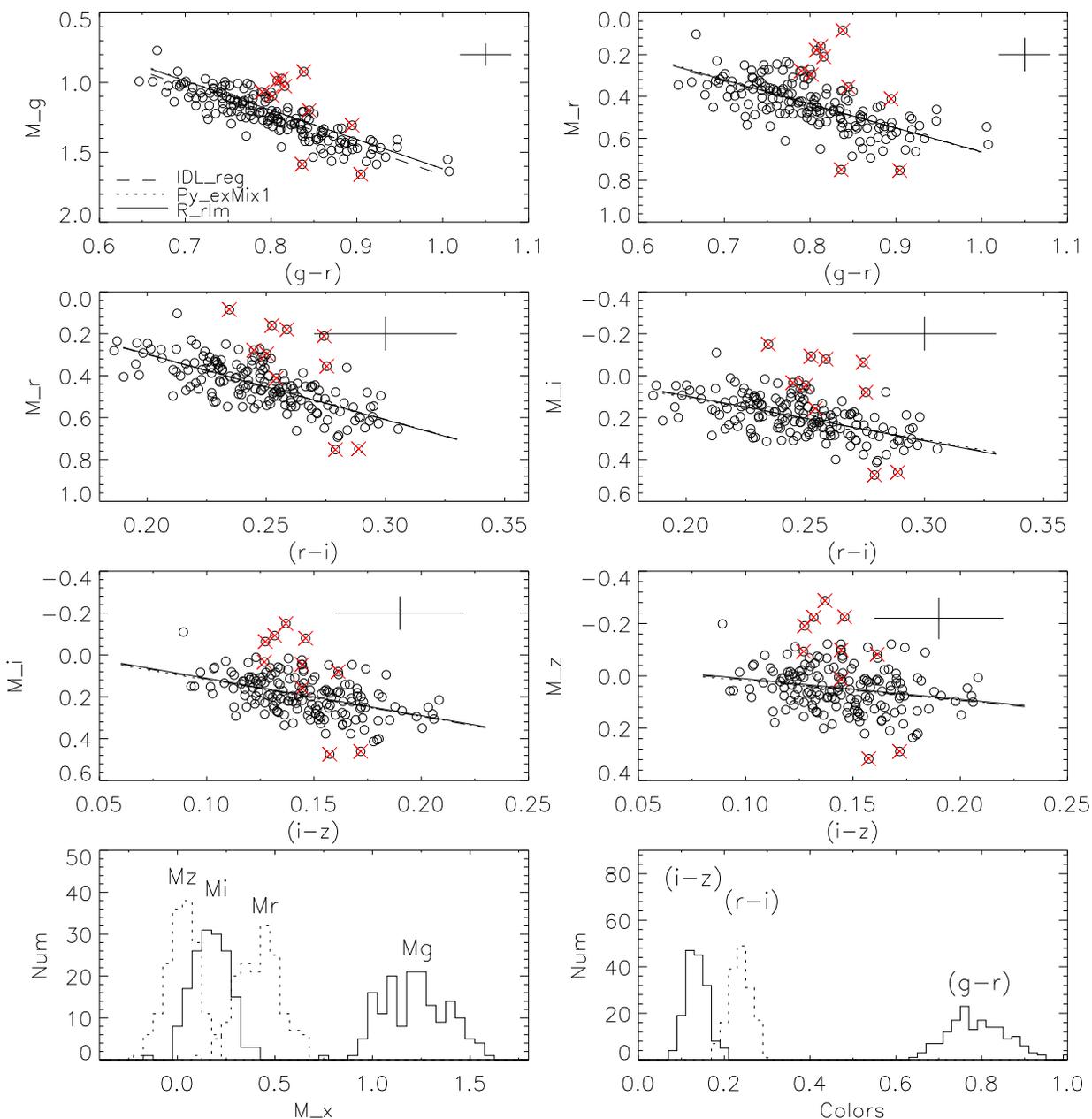}
\caption{Color and absolute magnitude diagrams and their
  distributions in the Sloan $griz$ system for our sample stars. Red crosses
  identify infrared anomalous stars as described in Section 3.4 and Figure 5.
  All colors and magnitudes are corrected for reddening.}
\end{figure*}

\subsection{2MASS}

We restrict ourself to stars with $JHK_s$ errors less than $0.03\,$ mag.
Fig.~5 shows
the $(J-H)$ versus $M_{J}$;  $(J-H)$ versus $M_{H}$; $(H-K_s)$ versus
$M_{H}$ and $(J-K_s)$ versus $M_{K_s}$ diagrams, as well as the 2MASS
color and
absolute magnitude distributions for our sample stars. There is a mild slope
in $M_{J}$ as a function of color $(J-H)$, while in the two
remaining filters $M_{H}$ and $M_{K_s}$ there is almost no trend with color.
This is quantified in Table 1 with the Pearson and Spearman correlation
coefficients.

In the widely used diagram of  $(J-K_s)$  versus  $M_{K_s}$, RC stars cover
the color range $0.5-0.7\,$mag, and most of them cluster at an absolute
magnitude that is consistent with the value $M_{K_s}=-1.613$ obtained by
\cite{Laney12} using local RC stars (indicated in the figure with a solid
line). Note that the value in \cite{Laney12} is already converted into
the 2MASS system, thus allowing direct comparison with our results.

\begin{figure*}
\epsscale{1.0}
\plotone{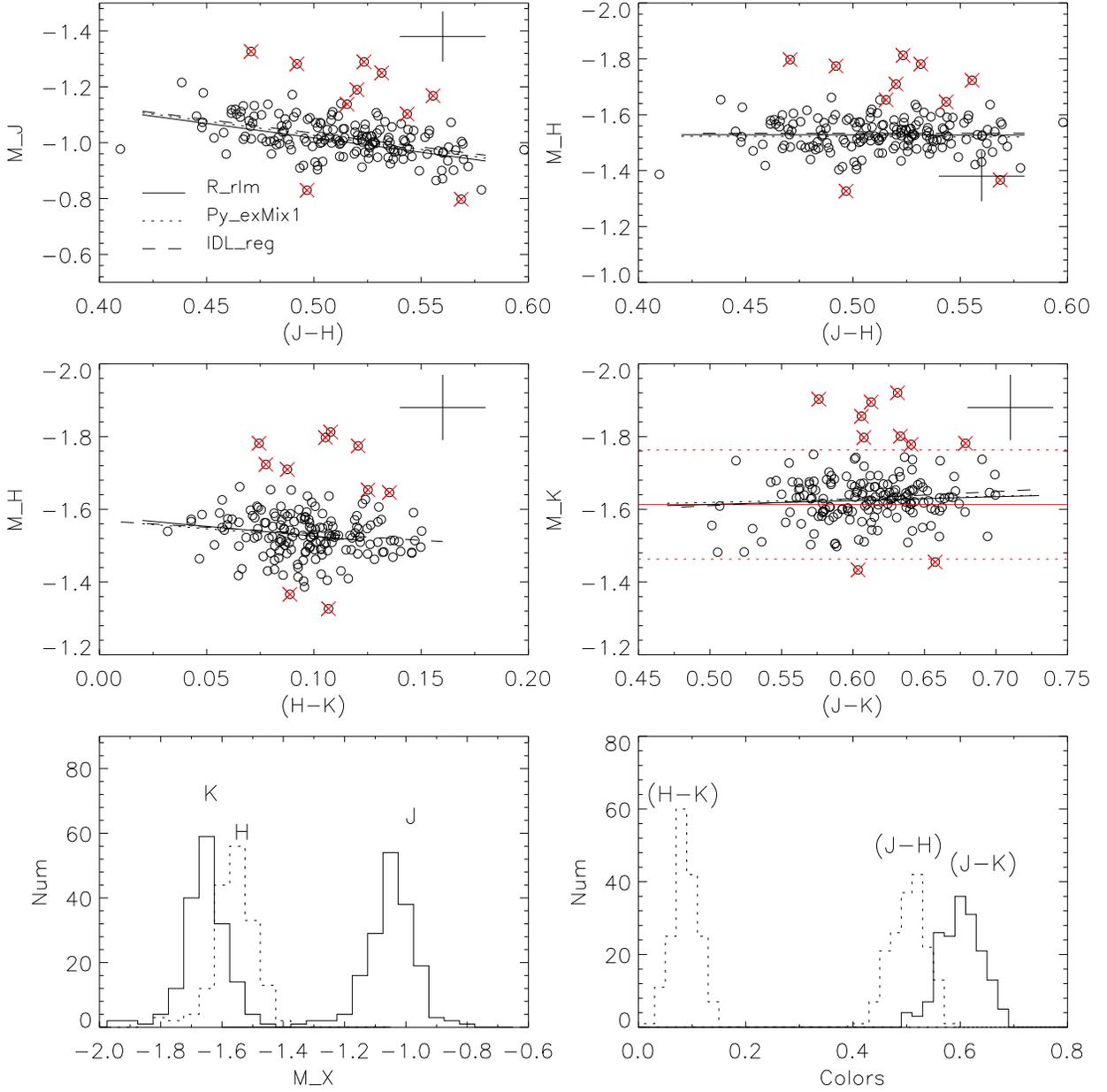}
\caption{Colour--absolute magnitude diagram of RC stars in various combination
  of 2MASS filters. The solid line in the $(J-K_s)$ versus $M_{K_s}$
  diagram is the absolute magnitude of local RC stars from \cite{Laney12}.
  Dashed lines indicate $\pm0.15\,$mag with respect to the absolute magnitude
  of local red clump stars, as explained in the text. Stars beyond the dashed
  lines are marked by red crosses. Bottom panels: absolute magnitude and color
  distributions for our sample of stars in the 2MASS system. All colors and
  magnitudes are corrected for reddening.}
\end{figure*}

For some stars, color and absolute magnitude combinations in the infrared
show significant deviations from the mean values of the whole sample, which is
not compatible with the maximum allowed photometric errors of $0.03\,$mag set
here, even after taking into account reddening and distance uncertainties on
absolute magnitudes.
Specifically, ten stars marked by red crosses in Fig.~5 lie beyond the dashed
lines, which represent a deviation of $0.15\,$mag (typical maximum error) from
the absolute magnitude $M_{K_s}=-1.613$ obtained by \cite{Laney12} from local
RC stars. The quoted 2MASS $JHK_s$ errors for the
these stars can not explain such large deviations. Most of these deviant stars
are overluminous, and this would point towards the presence of some sort of
infrared emission e.g., due to a hot circumstellar disk. Also, although we have
removed RC stars with ages below 2 Gyr from our analysis, five of the deviant
ones have
ages between 2 and 3 Gyr, and thus some residual age effect on absolute
magnitudes could still be present for some of these objects. Investigating
these scenarios however is beyond the scope of the present paper. We exclude
these stars in the rest of the analysis, and we mark them with red crosses
in the plots for reference.

\subsection{WISE}

Fig.~6 shows the color--absolute magnitude diagrams in the  WISE system.
Photometric uncertainties in $W1W2$ are of order $0.02$ mag, significantly
smaller than for $W3W4$, which are of order $0.02-0.10$ and $0.10-0.50$ mag,
respectively. Because of these uncertainties, we impose a threshold on the
maximum allowed photometric error: $0.03$ mag on $W1$ and $W2$, and $0.05$ mag
on $W3$, while we discard $W4$ from the rest of the analysis.

The color and magnitude ranges in the WISE system are quite small, which
indicates that these bands can be used to obtain very good distances
for RC stars. We quantify in Table 1 the Pearson and Spearman correlation
coefficients. Also, comparing Fig 4, 5 and 6 we see that the slope
of absolute magnitudes as function of colors flattens out, and then reverses
as moving to longer wavelength.

\begin{figure*}
\epsscale{1.0}
\plotone{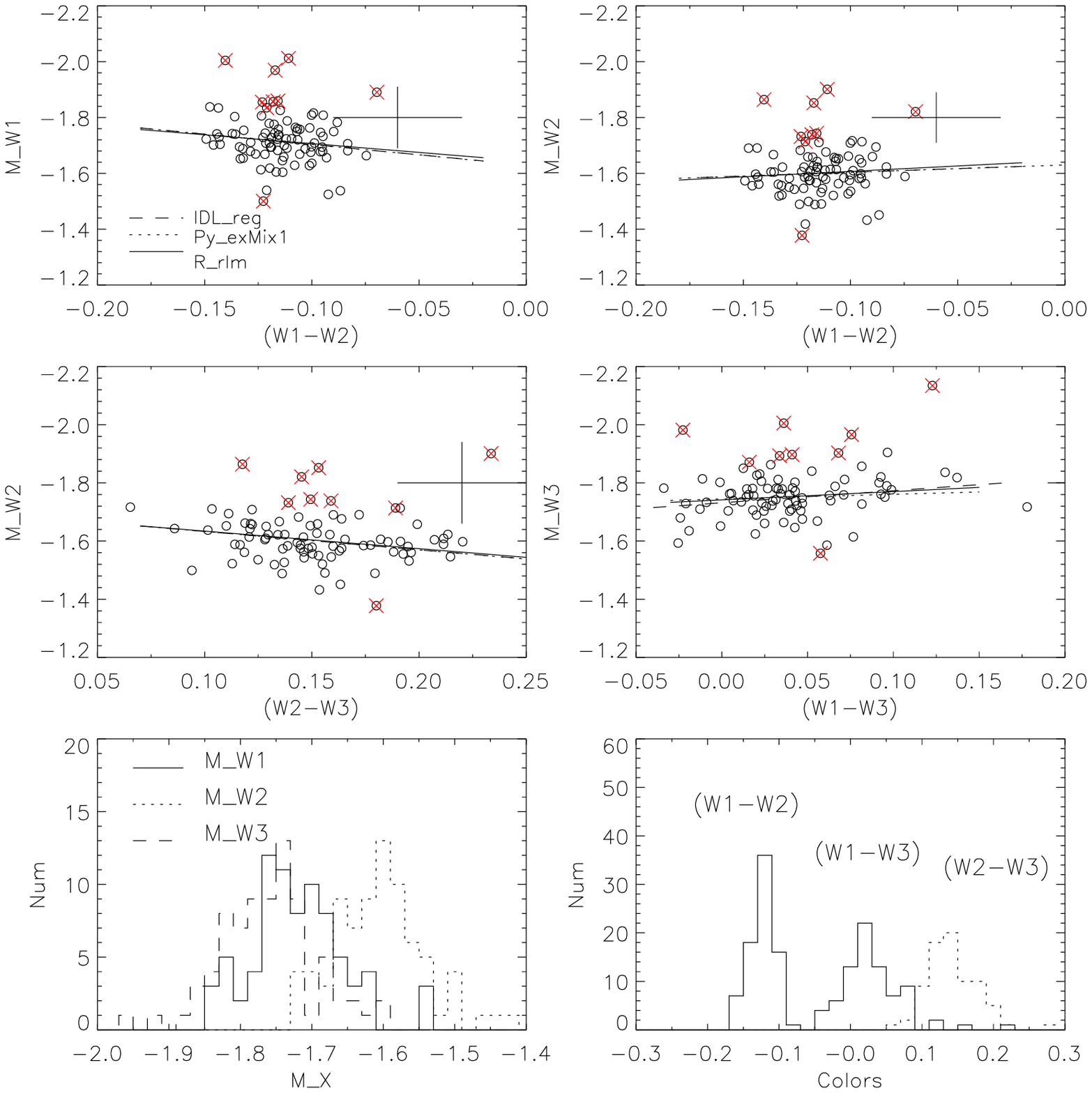}
\caption{Color--absolute magnitude diagrams in various combinations of WISE
  filters for our sample of stars. Red crosses same as in Fig.~5. All colors
  and magnitudes are corrected for reddening.}
\end{figure*}
\begin{landscape}
\begin{table*}
\caption{Mean colors and absolute magnitudes, together with the
color-absolute magnitude calibrations and Pearson/Spearman
correlation coefficients based on IDL-reg function, in different
 photometric systems for RC stars. Difference between mean and
 median values is usually a few millimag only, and never exceed
 $0.01\, mag$. We use the scatter of the data as a conservative
 estimate of the errors: if we were to use the standard deviation
 of the mean, uncertainties would be a factor of 10 smaller. Note
 that all colors and absolute magnitudes are corrected for reddening.}
\begin{tabular}{cccc}
\noalign{\smallskip}
\hline
$(b-y)  = 0.619 \pm 0.048$ & $M_y    =  0.754 \pm 0.147$ & $M_y=2.209 (b-y)-0.626$ & $0.72/0.72$\\
$(B-V)  = 1.025 \pm 0.140$ & $M_V    =  0.735 \pm 0.148$ & $M_V=0.189 (B-V)+0.525$ & $0.13/0.18$\\
$(g-r)  = 0.809 \pm 0.078$ & $M_g    =  1.229 \pm 0.172$ & $M_g=2.010 (g-r)-0.402$ & $0.88/0.89$\\
$(g-r)  = 0.809 \pm 0.078$ & $M_r    =  0.420 \pm 0.110$ & $M_r=1.010 (g-r)-0.402$ & $0.67/0.72$\\
$(r-i)  = 0.263 \pm 0.029$ & $M_r    =  0.420 \pm 0.110$ & $M_r=2.738 (r-i)-0.303$ & $0.65/0.71$\\
$(r-i)  = 0.263 \pm 0.029$ & $M_i    =  0.157 \pm 0.094$ & $M_i=1.738 (r-i)-0.303$ & $0.45/0.55$\\
$(i-z)  = 0.136 \pm 0.021$ & $M_i    =  0.157 \pm 0.094$ & $M_i=2.382 (i-z)-0.169$ & $0.53/0.52$\\
$(i-z)  = 0.136 \pm 0.021$ & $M_z    =  0.022 \pm 0.084$ & $M_z=1.382 (i-z)-0.169$ & $0.34/0.34$ \\
$(J-H)  = 0.513 \pm 0.034$ & $M_J    = -1.016 \pm 0.063$ & $M_J=0.975 (J-H)-1.518$ & $0.50/0.47$\\
$(H-K_s)= 0.097 \pm 0.024$ & $M_H    = -1.528 \pm 0.055$ & $M_H=0.494 (H-K_s)-1.580$ & $0.15/0.18$\\
$(J-K_s)= 0.609 \pm 0.040$ & $M_{K_s} = -1.626 \pm 0.057$ & $M_{K_s}=-0.188 (J-K_s)-1.517$ & $-0.17/-0.09$\\
$(y-K_s)= 2.383 \pm 0.129$ & $M_{K_s} = -1.626 \pm 0.057$ & $M_{K_s}=-0.003 (y-K_s)-1.625$ & $-0.03/0.08$\\
$(W1-W2)=-0.112 \pm 0.016$ & $M_{W1}  = -1.694 \pm 0.061$ & $M_{W1}=0.612 (W1-W2)-1.632$ & $0.17/0.13$\\
$(W2-W3)= 0.155 \pm 0.043$ & $M_{W2}  = -1.595 \pm 0.064$ &  $M_{W2}=0.628 (W2-W3)-1.696$ & $0.25/0.23$\\
$(W1-W3)= 0.043 \pm 0.044$ & $M_{W3}  = -1.752 \pm 0.068$ &  $M_{W3}=-0.307 (W1-W3)-1.741$ & $-0.36/-0.20$\\
$(H-W2)= 0.053 \pm 0.025$ & $M_{W2}  = -1.581 \pm 0.060$ &  $M_{W2}=-0.967 (H-W2)-1.533$ & $-0.41/-0.36$\\
\noalign{\smallskip} \hline
\end{tabular}
\end{table*}
\end{landscape}

\begin{table}
\caption{The color--absolute magnitude calibrations
  based on {\it R-rlm} function for RC stars. Note that all colors and absolute
  magnitudes are corrected for reddening.}
\begin{tabular}{c}
\noalign{\smallskip}
\hline
$M_y=2.206(\pm0.149)(b-y)-0.620 (\pm0.093)$ \\
$M_V=0.200\pm0.096)(B-V)+0.512 (\pm0.099)$\\
$M_g=2.099(\pm0.089)(g-r)-0.463 (\pm0.072)$\\
$M_r=1.109(\pm0.089)(g-r)-0.467 (\pm0.071)$\\
$M_r=3.117(\pm0.238)(r-i)-0.399 (\pm0.059)$\\
$M_i=2.116(\pm0.239)(r-i)-0.390 (\pm0.059)$\\
$M_i=1.752(\pm0.257)(i-z)-0.058 (\pm0.038)$\\
$M_z=0.753(\pm0.316)(i-z)-0.057 (\pm0.038)$\\
$M_J=1.011(\pm0.129)(J-H)-1.534 (\pm0.066)$\\
$M_H=0.536(\pm0.171)(H-K_s)-1.580 (\pm0.017)$\\
$M_{K_s}=-0.106(\pm0.104)(J-K_s)-1.560 (\pm0.063)$\\
$M_{K_s}=0.045(\pm0.031)(y-K_s)-1.740(\pm0.077)$\\
$M_{W1}=0.631(\pm0.426)(W1-W2)-1.643(\pm0.049)$\\
$M_{W2}=0.601(\pm0.175)(W2-W3)-1.694(\pm0.028)$\\
$M_{W3}=-0.289(\pm0.166)(W1-W3)-1.742(\pm0.010)$\\
$M_{W2}=-0.854(\pm0.197)(H-W2)-1.538(\pm0.011)$\\
\noalign{\smallskip} \hline
\end{tabular}
\end{table}

\subsection{The Combined Str\"omgren, 2MASS and WISE Systems.}

It is interesting to explore two widely used combination of color and absolute
magnitudes for RC stars. Namely, $(V-K_s)$
versus $M_{K_s}$ and $(H-W2)$ versus $M_{W2}$ are widely
adopted in the literature. Since $V$ magnitudes in APASS have somewhat large
errors, we adopt $y$ magnitudes in this analysis (where in fact, Str\"omgren
$y$ was historically defined to be essentially the same as Johnson $V$).
Fig.~7 shows the $(y-K_s)$ versus $M_{K_s}$, and the $(H-W2)$ versus
$M_{W2}$ diagrams, together with their color distributions.
There is essentially flat correlation between $(y-K_s)$ and $M_{K_s}$,
with a Pearson (Spearman) coefficient of $-0.03$ (0.08).
On the contrary, $M_{W2}$ inverse correlates with $(H-W2)$, the
Pearson (Spearman) coefficient being $-0.41$ ($-0.36$).
Color histograms are shown in the bottom of Fig.~7, and the mean values are
$(y-K_s)=2.383\pm0.129$ and $(H-W2)=0.053\pm0.025$. In particular, the
rather narrow range of $(H-W2)$ makes it a useful cryon to determine reddening
in high extinction areas using RC stars.

Our results based on {\it IDL-reg} function are summarized in Table 1.
Results from PYTHON {\it exMix1} and {\it R-rlm} are very close, and we
present the color--absolute magnitude calibrations based on {\it R-rlm}
function in Table 2.

\begin{figure*}
\epsscale{1.0}
\plotone{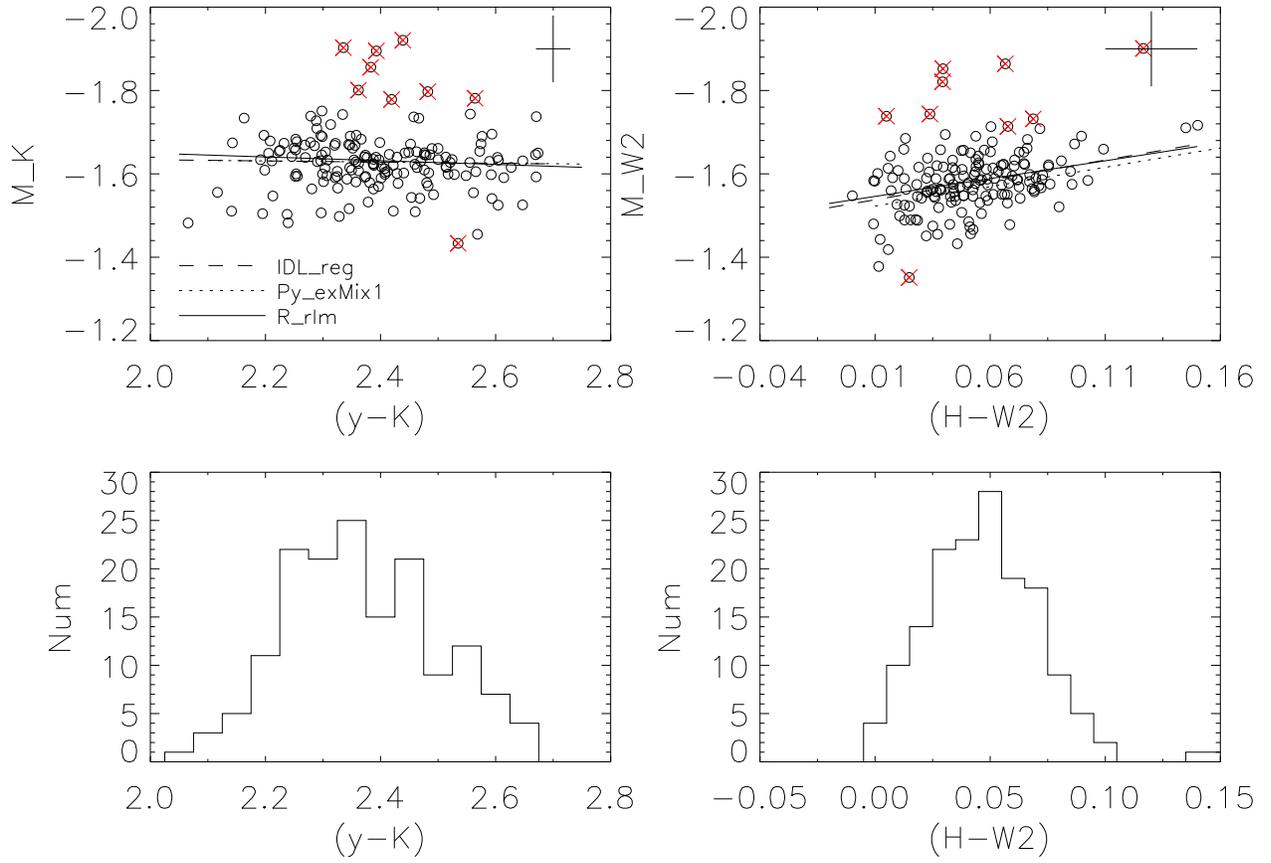}
\caption{Color--absolute magnitude diagram for RC in the $(y-K_s)$ versus
  $M_{K_s}$ and $(H-W2)$ versus $M_{W2}$ system. Red crosses are the same as in
  Fig.~5. Bottom panels show the color distributions. All colors and
  magnitudes are corrected for reddening.}
\end{figure*}

\subsection{The age dependence of the distance scale.}

Using the age (and mass) information available from SAGA, in Fig.~8 we
plot absolute magnitudes of RC stars as function of age, in the $K_s$ and $W2$
bands, which are two among the filters displaying the least colour-dependence.
Stars with age errors larger than 30\% are removed from this plot,
although they would still follow the same trend if included.
Interestingly, older than $2$~Gyr there is a clear dependence of $K_s$ and $W2$
absolute magnitudes on age.
Fitting this trend with {\it IDL-reg}, we obtain:
\begin{equation}
M_{K_s}=(0.015\pm0.003)\,\tau - 1.715(\pm0.016)
\end{equation}
and
\begin{equation}
M_{W2}=(0.017\pm0.003)\,\tau -1.682(\pm0.016)
\end{equation}
where $\tau$ is the age of RC stars in Gyr. Fits using {\it R-rlm} and
{\it Py-exMix1} are
very similar.  $M_{K_s} = (0.016\pm0.002)\,\tau -1.715(\pm0.008)$,
$M_{W2} = (0.018\pm0.002)\,\tau -1.683$ with the former method and
$M_{K_s} = (0.015\pm0.003)\,\tau - 1.714$ and
$M_{W2}=(0.017\pm0.004)\,\tau -1.682$ with the latter.
The Pearsons (Spearman) correlation coefficients are of 0.65 (0.62) for
$M_{K_s}$
and 0.66 (0.63) for $M_{W2}$, indicating a strong linear correlations. In
Table 3 we report the slope of ages versus magnitudes for all our
photometric systems. It has to be kept in mind that certain filters also
display color dependence, and thus the interdependence of ages and colors
might not be straightforward to disentangle. In general, we can say that for
optical colors there is a slope of $\sim 0.030 \pm 0.003\,\rm{mag\,Gyr}^{-1}$
whereas in the infrared the dependence is
$\sim 0.020 \pm 0.003\,\rm{mag\,Gyr}^{-1}$ and it also displays a stronger
correlation.

Note that the asteroseismic ages adopted here are obtained assuming no
mass-loss. If we were to use instead asteroseismic ages derived assuming a
highly efficient mass-loss, the slopes above would increase even further.
E.g., the slopes of $M_{K_s}$ and $M_{W2}$ would be
$\sim 0.031\pm0.011\,\rm{mag\,Gyr}^{-1}$ and
$\sim0.035\pm0.011\,\rm{mag\,Gyr}^{-1}$. We refer to
\cite{C16} for a discussion why a negligible mass-loss is however favored by
current observations.

The dependence of absolute magnitudes of RC stars on age is theoretically
predicted \citep[see][in $I$ band]{Girardi01} and found in open clusters by
\cite{Grocholski02} using 2MASS photometry. Here, for the first time, we
detect such a tiny trend in various optical and infrared bands in field stars
thanks to the accuracy of our asteroseismic ages and distances (and hence
luminosities).

\begin{figure*}
\epsscale{1.0}
\plotone{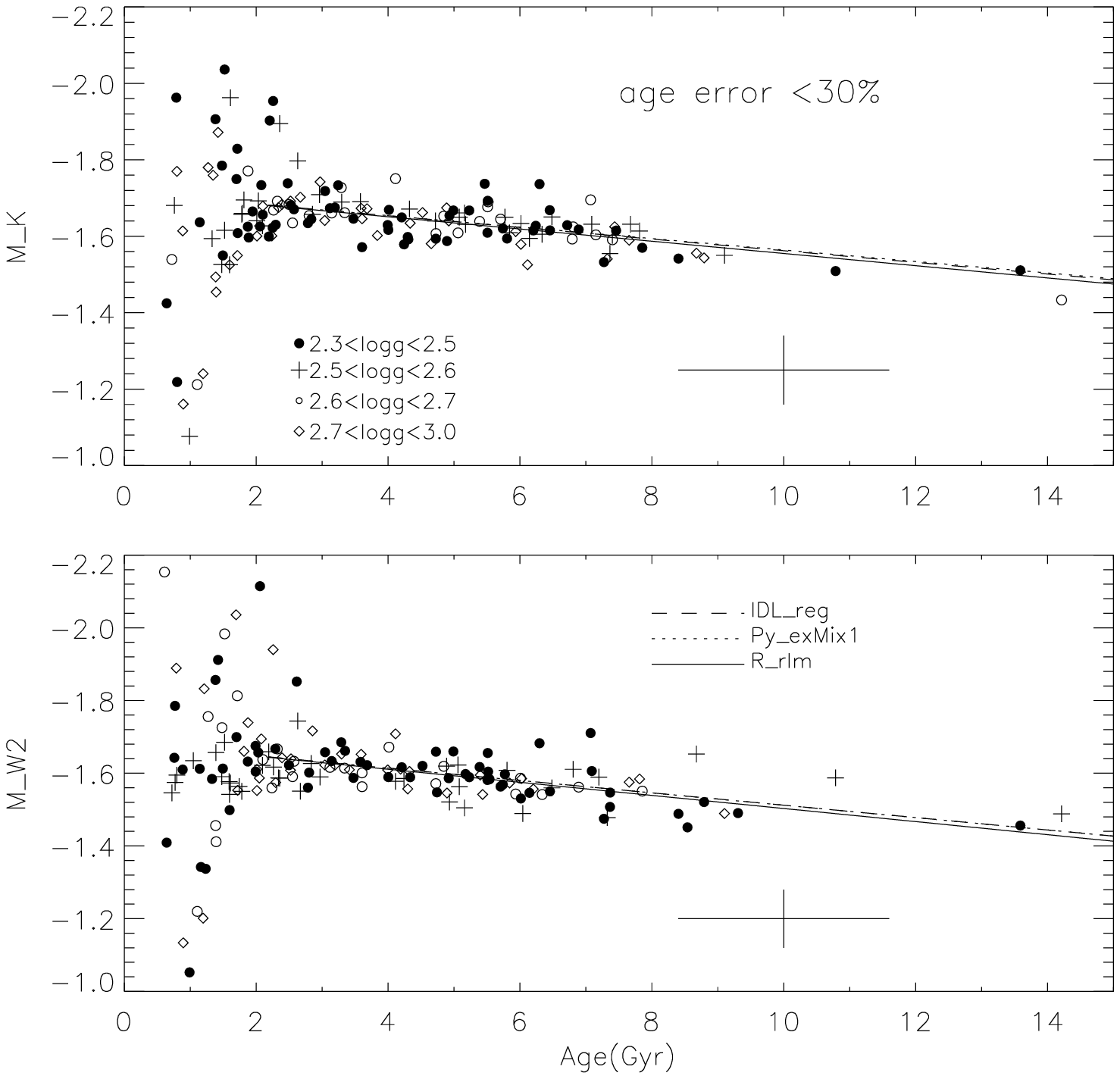}
\caption{$M_{K_s}$ and $M_{W2}$ as function of stellar ages for different
  $\logg$ ranges. Absolute magnitudes are corrected for reddening. Ages below
  2 Gyr are plotted only to highlight their wide range of absolute magnitudes,
  but are not used for the fits. Stars with age error larger than 30\% are not
  shown.}
\end{figure*}

\begin{table}
  \caption{Slope of absolute magnitude versus ages in $\rm{mag\,Gyr}^{-1}$ for
different filters. Last column indicates Pearson and Spearman correlation
coefficients.}
\begin{tabular}{ccc}
\noalign{\smallskip}
\hline
band & slope & correlation \\
\hline
$y$   & $0.032\pm0.003$ & $0.43/0.43$ \\
$V$   & $0.028\pm0.003$ & $0.40/0.41$ \\
$g$   & $0.036\pm0.003$ & $0.40/0.42$ \\
$r$   & $0.029\pm0.003$ & $0.52/0.50$ \\
$i$   & $0.026\pm0.003$ & $0.55/0.53$ \\
$z$   & $0.025\pm0.003$ & $0.61/0.60$ \\
$J$   & $0.020\pm0.003$ & $0.63/0.62$ \\
$H$   & $0.018\pm0.003$ & $0.71/0.69$ \\
$K_s$ & $0.015\pm0.003$ & $0.65/0.62$ \\
$W1$  & $0.016\pm0.003$ & $0.62/0.59$ \\
$W2$  & $0.017\pm0.003$ & $0.66/0.63$ \\
$W3$  & $0.017\pm0.004$ & $0.51/0.48$ \\
\noalign{\smallskip} \hline
\end{tabular}
\end{table}

\section{Literature comparison: using the distance scale to test asteroseismic
  scaling relations}

In the Johnson system, $M_V$ are generally consistent with
previous results from \citet{Bilir13a} based on red clump/red horizontal branch
stars in globular and open clusters. They also found a weak metallicity
dependence of the $M_V$ with a coefficient of $0.046$ added to the magnitude
versus $(B-V)$ calibration.
The weak dependence of $M_y$ on $\feh$ in our work in the Str\"omgren system
is consistent with the result of \citet{Bilir13a}.
The absolute magnitudes of RC stars are calibrated in \cite{Bilir13b} in terms
of colors for $M_V$, $M_J$, $M_{K_s}$ and $M_g$.
Our color--absolute magnitude diagrams are consistent with those in
\cite{Bilir13b} (their figure 8), although our seismic selection of RC stars
results in a smaller scatter. Our mean values of $M_J=-1.016\pm0.063$ and
$M_H=-1.528\pm0.055$ agree within the errors with those of \cite{Laney12} who
gave
$M_J=-0.984\pm0.014$ and $M_H=-1.490\pm0.015$ based on local RC stars.

The comparison with $M_{K_s}$ values from the literature indicates an overall
agreement, although it deserves a discussion. Our value
of $M_{K_s}=-1.626\pm0.057$ is consistent with that of $-1.613\pm0.015$ by
\cite{Laney12} for local RC stars, and that of $-1.61\pm0.04$ by
\cite{Grocholski02} for RC stars in clusters. The values of
$-1.57\pm0.05\,$mag by \cite{Helshoecht07} and of  $-1.54\pm0.04\,$mag by
\cite{Groenewegen08} for local RC stars with {\it Hipparcos} parallaxes
are fainter than our value.
The comparison of our values of $M_{W1}=-1.694\pm0.061$
and $M_{W3}=-1.752\pm0.068$ with those of $M_{W1}=-1.635\pm0.026$
and $M_{W3}=-1.606\pm0.024$ in \cite{Yaz13} shows deviations
of $-0.06\,$mag in $M_{W1}$ and $-0.15\,$mag in $M_{W3}$ (ours minus theirs,
our absolute magnitudes being brighter). \cite{Yaz13} also provide
$M_{J}$, $M_{H}$
and $M_{K_s}$, and again our absolute magnitudes are brighter by $\simeq -0.03$
to $-0.06$ mag. From the above comparisons, we conclude that differences with
the {\it Hipparcos} literature are generally within the errors, although it
is intriguing to notice that they seem systematic, in the sense that our
absolute magnitudes are brighter.

Our absolute magnitudes are based on seismic distances $D$, which are
derived scaling angular diameters obtained via the InfraRed Flux Method to the
stellar radii $R$ obtained from scaling relations \citep{Silva12,C14a}. We
have carried out an extensive comparison of our
angular diameters with interferometric measurements in \cite{C14b}. If we
assume our angular diameter scale to be correct, then any change in stellar
radii due to scaling relation directly translates into a change of distances.
Thus, we can compare the absolute magnitude of clump stars we derive (i.e.
based on distances relying on scaling relations) with those available in the
literature and obtained with independent methods. From distance modulus, it
follows that for a given photometric system a difference $\Delta M=M-M_l$ in
absolute magnitudes between our calibrations ($M$) and those available in the
literature $M_l$ corresponds to a fractional change in distance
\begin{equation}
  \frac{D}{D_l}=10^{-0.2 \Delta M}
\end{equation}
which can then be used to constrain how much seismic radii $R$ should vary to
agree with the distance scale in the literature. In this way, we can thus put
an upper limit to the precision of scaling relations.

For this purpose, we compare the difference of the absolute magnitudes with
those in the literature for $M_{K_s}$, the filter showing the least dependence
on colors, and only mildly affected by reddening. The offsets are
$-0.013$\,mag with respect to the absolute magnitude
in \cite{Laney12}, $-0.056$\,mag with respect to \cite{Helshoecht07},
$-0.086$\,mag with respect to \cite{Groenewegen08} and $-0.016$\,mag with
respect to \cite{Grocholski02}.
This corresponds to our seismic distances being $10^{-0.2 \Delta M}$, i.e.
$1.006$, $1.026$, $1.040$ and $1.007$ larger. If we assume that radii from
scaling relation are entirely responsible for the difference, then radii from
scaling relations are overestimated by an amount that goes from
$0.6\pm 2.7\%$ to $4.0\pm 3.2\%$ depending on the literature calibration we
compare with. The average offset is $2\pm2\%$. This comparison indicates a
mild tension between our seismic distance scale, and that deduced from
{\it Hipparcos} parallaxes. If confirmed, it would put a limit to the accuracy
at which scaling relations are applicable to red clump stars. However, a few
caveats must be remembered. Our comparison
assumes no color dependence (i.e. we only compare mean absolute
magnitudes). Also, one may wonder whether the age distribution underlying our
sample is different from that of other calibrations. Since all calibrations
are based on
nearby stars, or open clusters, it is reasonable to assume that the underlying
age distributions are comparable. However, calibrations using field stars in
the literature have no age information, and likely include many stars younger
than 2~Gyr, which instead we have excluded. If we were to include stars of all
ages in our calibration, the mean absolute magnitude for our sample
would be $M_{K_s}=-1.628\pm0.130$, i.e. the difference would stay the same,
but the scatter would increase significantly.

\section{Summary}
Based on the reddening and distance estimates for a sample of $\sim170$
seismically identified red clump stars in \cite{C14a}, we have
investigated color and absolute magnitude distributions of RC stars
in Str\"omgren $by$, Johnson $BV$, Sloan $griz$, 2MASS $JHK_s$ and WISE $W1W2W3$
photometric systems. For the first time, we find a clear trend between absolute
magnitudes and ages in field RC stars. The absolute magnitudes of RC stars
deviate significantly from a constant value at ages below 2
Gyr, which indicates that RC stars can be reliably used as distance indicators
only for populations older than this age. Even so, a statistically significant
correlation between absolute magnitudes and ages remain, which in worst
case can introduce a bias up to $\sim0.3\,$mag in the optical and $\sim0.2\,$mag
in the infrared if ages of clump stars are not known. Our absolute magnitudes
for RC
stars in the 2MASS and WISE system are generally consistent within the errors
with those obtained from local RC stars with accurate {\it Hipparcos}
parallaxes. However, a possible tension at the level of a few percent is
identified. Assuming that
seismic scaling relations are responsible for this difference, this would imply
that seismic radii for red clump stars are overestimated by $2\pm2\%$ when
using scaling relations. Our methodology, along with improvements on the
calibration of the distance scale with future Gaia data releases will be able
to shed light on this issue.

\acknowledgments
This study is supported by the National Natural Science Foundation of China
under grant Nos. 11625313, 11573035, 11233004, 11390371, 11052037, the
National Scholarship Council Fund No. CSC201604910337 and the National Key
Basic Research Program of China (973 program) Nos. 2014CB845703.
L.C gratefully acknowledge support from the Australian Research Council
(grants DP150100250, FT160100402). L.C. acknowledge the generous hospitality
of the National Astronomical Observatory of the Chinese Academy of Sciences,
where this work was initiated.
Funding for the Stellar Astrophysics Centre is provided by The Danish National
Research Foundation (Grant DNRF106). V.S.A. acknowledges support from VILLUM
FONDEN (research grant 10118)

\end{document}